\documentclass[sigconf]{acmart}
\usepackage{amsthm}
\usepackage{amsmath}
\usepackage{algorithm}
\usepackage{algpseudocode}

\AtBeginDocument{%
  }

\theoremstyle{remark}
\newtheorem{problem}{Problem}

\copyrightyear{2023}
\acmYear{2023}
\setcopyright{acmlicensed}\acmConference[MSWiM '23]{Proceedings of the Int'l ACM Conference on Modeling Analysis and Simulation of Wireless and Mobile Systems}{October 30-November 3, 2023}{Montreal, QC, Canada}
\acmBooktitle{Proceedings of the Int'l ACM Conference on Modeling Analysis and Simulation of Wireless and Mobile Systems (MSWiM '23), October 30-November 3, 2023, Montreal, QC, Canada}
\acmPrice{15.00}
\acmDOI{10.1145/3616388.3617550}
\acmISBN{979-8-4007-0366-9/23/10}

\begin{document}

\title{High Fidelity Fast Simulation of Human in the Loop Human in the Plant (HIL-HIP) Systems}

 \author{Ayan Banerjee, Payal Kamboj, Aranyak Maity, Riya Sudhakar Salian, Sandeep K.S. Gupta}
 \email{{abanerj3,pkamboj,amaity1,rsalian1,sandeep.gupta}@asu.edu}
 \affiliation{%
   \institution{School of Computing and Augmented Intelligence, Arizona State University}
   \state{Arizona}
   \country{USA}
 }

\begin{abstract}
   Non-linearities in simulation arise from the time variance in wireless mobile networks when integrated with human in the loop, human in the plant (HIL-HIP) physical systems under dynamic contexts, leading to simulation slowdown. Time variance is handled by deriving a series of piece wise linear time invariant simulations (PLIS) in intervals, which are then concatenated in time domain. In this paper, we conduct a formal analysis of the impact of discretizing time-varying components in wireless network-controlled HIL-HIP systems on simulation accuracy and speedup, and evaluate trade-offs with reliable guarantees. We develop an accurate simulation framework for an artificial pancreas wireless network system that controls blood glucose in Type 1 Diabetes patients with time varying properties such as physiological changes associated with psychological stress and meal patterns. PLIS approach achieves accurate simulation with $>2.1$ times speedup than a non-linear system simulation for the given dataset.
\end{abstract}





\maketitle

\section{Introduction}
\begin{figure}
\center
\includegraphics[trim=0 0 0 0,width=\columnwidth]{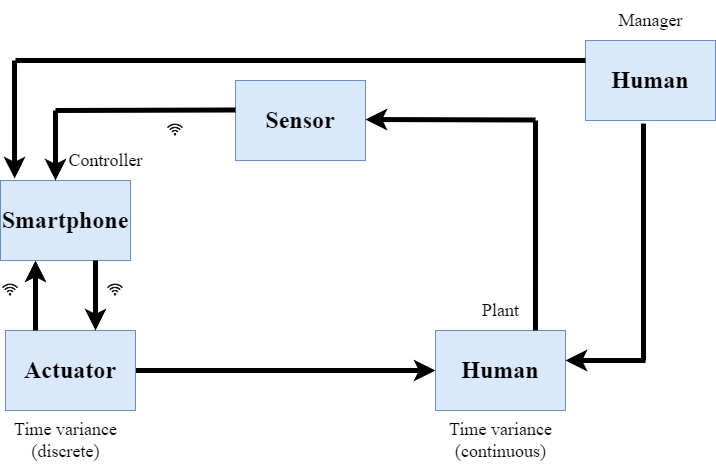} 
 \caption{The HIL-HIP system model for a wireless network controlled plant where the human is both the configuration manager and a component of the plant.}
    \label{fig:HilHip}
\end{figure}

Modern day safety critical human in the plant (HIP) infrastructure such as artificial pancreas, and autonomous cars, work through integration of a wireless network with a physical plant often including a human for sensing and actuation (Fig. \ref{fig:HilHip}). Irrespective of autonomy level, a human manager, human in the loop (HIL), often makes configuration changes to achieve the best performance. The simulation of the integration of a wireless mobile network (WMN) driven control system with the HIL-HIP architecture is essential for performance, safety, and resource efficacy analysis.    

 The tight coupling of the WMN with the human manager/user results in frequent changes in control configurations as well as changes in the physical system properties~\cite{Banerjee15TMC,Banerjee12Your}. Consequently, the WMN exhibits time-varying characteristics. The time varying properties of the discrete control software executed by the WMN changes in-frequently and can be tackled using an event driven simulation strategy. As such simulation of time varying physical dynamics through an event driven simulation strategy will be prohibitively expensive since it will result in a simulation time step that is infinitesimally small. 

\begin{table*}
  \caption{Related Works}
  \label{Related}
  \centering
  \scriptsize
  \begin{tabular}{p{1.2cm} p{3.3cm} p{3.8cm} p{2.1cm} p{5.0cm}}
    \toprule         
    Study & System & Method & Accuracy Guarantees & Metrics for Results  \\
    \midrule
     \cite{choi2020can}
    & Heat exchange ventilation system 
    & Reduced order polynomial ODE 
    & No 
    & Low RMSE and error rate. \\
    \cite{hartmann2013progress}
    &Recharge areas of Karst system 
    & Reduced order polynomial ODE 
    &No 
    &Reproduction of recharge areas of the Karst system \\
    \cite{ning2022euler} 
    &Compartmental Model 
    &Physics Informed Neural Networks.
    &No
    &Simulation accuracy \\
    \cite{dong2008general} 
    &A nonlinear system
    & Piecewise-polynomial representations.
    &No
    &Average speedup w.r.t full model.\\
    \cite{michalak2017development} 
    &Ventilation flow system
    & ODE-45 solver 
    &No
    &Simulation accuracy\\\midrule
    \textbf{This work} & \textbf{Artificial Pancreas} & \textbf{piecewise linear time invariant} & \textbf{Yes} & \textbf{Speed up accuracy tradeoff} \\    
    \bottomrule
  \end{tabular}
\end{table*}

 Traditionally, a time varying system is treated as a non-linear system, and complex non-linear solvers are used for simulation which may require significant execution time~\cite{Lamrani21Operational}. This prevents their usage in applications such as digital twins~\cite{Nabar11GEM}, forward safety analysis~\cite{Banerjee13Stha} or online observer~\cite{banerjee2023statistical,maity2022cyphytest}. An obvious simplification is modeling the time varying system as a concatenation of several piecewise time invariant systems. The simulation time interval $[0, T]$ is subdivided into sub-intervals starting at times $\tau_j$. At the start of each time interval, a zero order hold assumption of the time varying parameters is undertaken, and the simulation is performed using linear system solution techniques such as Euler method~\cite{jameson1983solution}. Although this enables the usage of simpler solvers and hence saves time, to the best of our knowledge there is no work that bounds the error rate of such piecewise time invariant approximations.

In this paper, we develop a theoretical framework to evaluate the simulation error of piecewise linear time invariant simulation (PLIS) approach. We derive a closed form solution for the propagation of error in model coefficients through the system dynamics and provide a pathway to derive time stamps of each simulation piece such that the simulation error can be kept within pre-specified bounds. We compare the PLIS simulator with two other types of simulation: a) ORACLE, which considered the non-linear system in presence of time variant dynamics, and b) Koopman simulator, that approximates the non-linear dynamical system as a higher order linear system. We show the execution of PLIS on three different control approaches using WMN for the artificial pancreas case study. PLIS can be configured to achieve the similar simulation accuracy as ORACLE or Koopman with at least 2.1 and at most 8 times speedup.

\section{Related Works}
Time-varying system dynamics can be simulated by solving nonlinear ordinary differential equations (ODE) \cite{michalak2017development}. Recent advancements take advantage of data driven supervised learning, such as NeuralODE,\cite{Chen18NODE} or Euler-physics induced neural networks~\cite{ning2022euler}. Such methods are usually slow and depending on the simulation time step. Moreover, no accuracy guarantees are provided.

Time varying systems can be represented with reduced-order polynomial models of the system which are then used to simulate the essential characteristics of the system and have demonstrated considerable improvement over traditional ODE solvers showing 6 - 9 times speed up in simulating electrical circuits~\cite{dong2008general}, and applied to ventillaotr simulation~\cite{choi2020can} and hydrological Karst model~\cite{hartmann2013progress}. However, these works are application specific and do not provide a closed form relation of speed-up with accuracy. 

In contrast, this paper shows a general time varying system simulation framework with closed form speedup and accuracy tradeoffs from which error bounds can be derived.

\section{Simulation Approach}

The configuration of the WMN (Fig \ref{fig:HilHip}) is given by a set of $p$ state variables $Y \in \mathbf{Z}^p$, $\mathbf{Z}$ is the set of natural numbers. The configurations $Y(t)$ are a function of time. The plant dynamics is represented using the set of $n$ state variables $X \in \mathbf{R}^n$, where $\mathbf{R}$ is the set of real numbers. The dynamics are time varying given by -
\begin{equation}
    \label{eqn:state}
    \scriptsize
    \frac{dX(t)}{dt} = \mathbf{A}(t) X(t) + \mathbf{B}(t) u(t),
\end{equation}
where $\mathbf{A}(t)$ is $n \times n$ and $\mathbf{B}(t)$ is $n \times m$ time varying parameters of the system and $u(t)$ is a $m \times  1$ vector of inputs obtained from the WMN. $u(t) = g(Y,X)$ is a function $g(.,.): (\mathbf{Z}^p, \mathbf{R}^n) \rightarrow \mathbf{R}^m$ of the WMN configuration $Y$ and the current plant state $X$ to the $m$ dimensional real space representing the computing algorithm. 

\noindent{\bf Input modeling:} In this manuscript we consider the step input i.e., $u(t) = a: a \in \mathbf{R}, if$ $t > 0, else$ $ u(t) = 0$. The most common type of inputs are square wave inputs. We assume that the wave width of the square wave will be larger than a simulation time step. Without loss in generality we can then assume $u(t)$ to be a step function since our analysis is limited to the time window when all step inputs have already executed their leading edge.

A trajectory $\zeta$ is a function from a set $[0, T]$, $T \in \mathbf{R}^{\geq 0}$, denoting time and WMN configurations $Y(t)$ to a compact set of values $\in \mathcal{R}$. Each trajectory is the output of the physical system model $M$ in the form of Eqn. \ref{eqn:state} for a WMN configuration $Y$ and the algorithm $g(.,.)$. Concatenation of $q$ output trajectories over time $\zeta(Y(t_0),t_1-t_0) \zeta(Y(t_1),t_2-t_1) \ldots \zeta(Y(t_{q-1}),t_q-t_{q-1})$ is a trace $\mathcal{T}$.      


\begin{definition}{ORACLE Simulator:} The oracle simulator $\Sigma_o$ has access to: a) closed form function $A_{i,j}(t) = f^A_{i,j}(t)$, and $B_{i,j}(t) = f^B_{i,j}(t)$ for each time varying model coefficient of the plant, and b) all WMN configuration changes of the future. It uses non-linear ODE solvers to evaluate the traces. 
\end{definition}

\begin{definition}{Koopman Simulator:} The Koopman simulator $\Sigma_k$ has access to a high dimensional linear time invariant system, referred to as Koopman LTI,  -
\begin{equation}
\scriptsize
    \frac{dX_k(t)}{dt} = A_k X_k(t) + B_k u(t), 
\end{equation}
where $A_k$ ($n_k \times n_k$) and $B_k$ ($n_k \times m$) are constants, where $n_k >> n$, and $X_k(t) = M_f(X(t))$ is a measurement function. The simulator obtains an estimate of $X_e(t)$ by computing the pre-image $M_f^{-1}$ of the measurement function. The Koopman LTI is developed so that $dist_e(X_e(t),X(t))<\epsilon_k$, where $\epsilon_k$ is the error of the Koopman transformation operation. In our work, we have utilized the discrete mode decomposition (DMD) method to obtain the Koopman transforms of the non-linear Eqn. \ref{eqn:state}~\cite{williams2015data}.   
\end{definition}

\begin{definition}{Piece-wise Linear-Invariant Simulator (PLIS)} divides the time interval $[0, T]$ into sub-intervals $\tau_j$, $j \in \{1 \ldots s\}$ with $\tau_s = T, \tau_j < \tau_{j+1} \forall j$, such that in a time interval $\tau_j$ the plant follows -
\begin{equation}
\label{eqn:stateP}
\scriptsize
    \frac{dX_{P}(t)}{dt} = A^j X_{P}(t) + B^j u(t),
\end{equation}
where $A^j = f_a(A(t),\tau_j)$ ($B^j = f_a(B(t),\tau_j)$), is zero order hold approximation of $A(t)$ ($B(t)$) in the time interval $\tau_j$, and $X_P$ is an estimated state vector.
\end{definition}

\begin{problem}{\bf PLIS design Problem:} Given a trace $\mathcal{T}$, find a PLIS $\Sigma_P = \{\tau_j,A^j,B^j\} : (\tau_s = T) \bigcap \tau_j < \tau_{j+1} \forall j \bigcap $  $dist(\zeta_p(Y(\tau_j)$ $,\tau_{j+1}-\tau_j),\zeta(Y(\tau_j),\tau_{j+1}-\tau_j)) < \epsilon_p \bigcap dist(\mathcal{T}_P,\mathcal{T}) < \psi_p$, with error $\epsilon_p$ in trajectory and $\psi_p$ in trace. Here $\zeta_p$ and $\mathcal{T}_P$ are the trajectories and traces for the PLIS.    
\end{problem}

This is a dual error analysis approach. We not only minimize the error in trajectory between two events, but also minimize the error for the entire trace including multiple WMN control events. 

\begin{figure}
\center
\includegraphics[trim=0 0 0 0,width=0.8\columnwidth]{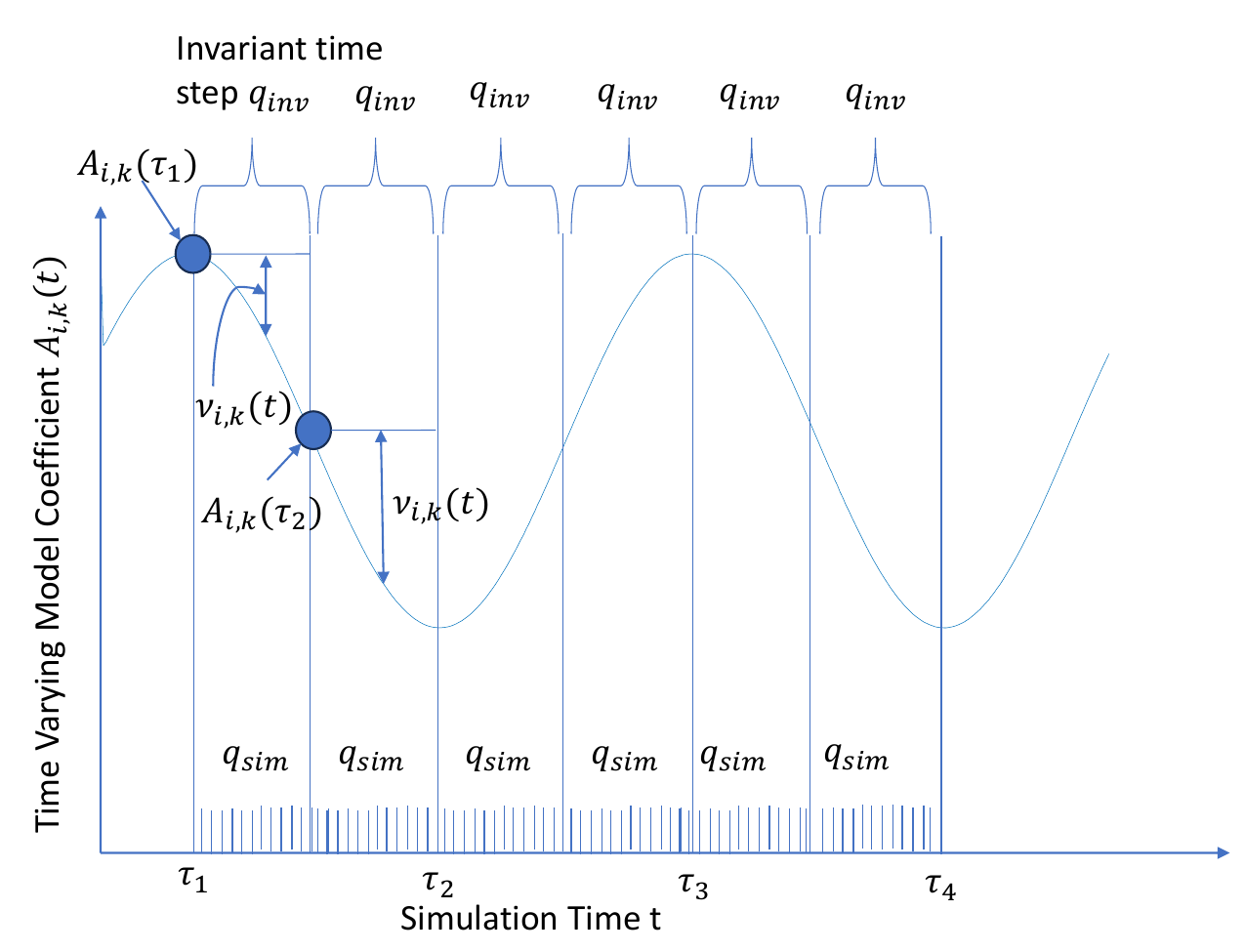} 
 \caption{PLIS simulation strategy to tackle non-linearities induced by time variance of model coefficients. The zero order hold strategy assumes constant value for $A_{i,k}(t)$ at each time sub-interval.}
    \label{fig:SimEx}
    \vspace{-0.1 in}
\end{figure} 

\noindent{\bf PLIS Design Solution:} For solving PLIS we need to find the function $f_a$, i.e., the zero order hold approximations of the coefficients. We assume that $f_a(A(t),\tau_j) = A(\tau_j), \forall t \in \{\tau_j,\tau_{j+1}\}$ and $f_a(B(t),\tau_j) = B(\tau_j), \forall t \in \{\tau_j,\tau_{j+1}\}$ (Fig. \ref{fig:SimEx}). The PLIS simulation divides the time interval $[0, T]$ into sub-intervals $\tau_j$ ($q_{inv}$ shown in Fig. \ref{fig:SimEx}) such that the propagated error from the coefficients to the trajectories is less than $\epsilon_p$, and to the overall trace is less than $\psi_p$. To this effect, we define two time steps: a) simulation time step, $q_{sim}$, is the fine grain division of each sub-interval so that the assumed time invariant dynamics in each sub-interval can be computed using the Euler method limiting the error within tolerance levels, and b) Invariant time step $q_{inv}$, (Fig. \ref{fig:SimEx}) determines the sub-intervals where the time invariance assumption results in a simulation error that is less than $\epsilon_p$ for every trajectory, and less than $\psi_p$ for the entire trace, i.e., $\tau_{j+1}-\tau_j = q_{inv} \forall j$. 
 
\noindent{\bf Error propagation:} For any element $A_{i,k} \in A$ ($B_{i,k} \in B$), the true value can be represented using an error value $A_{i,k}(t) = A_{i,k}(\tau_j) + \nu_{i,k}(t)$. From Eqn. \ref{eqn:state},  we obtain the following error propagation model for an element $x^i_p \in X_P$ -
\begin{equation}
    \label{eqn:Prop}
    \scriptsize
    \frac{dx^i_p}{dt} = \sum^{n}_{k=1}{A_{i,k}(\tau_j) x^k_p} + \sum_{k = 1}^{m}{B_{i,k}(\tau_j)u^k(t)} + \sum^{n}_{k=1}{A_{i,k}\mu_{k}} + \frac{d\mu_i}{dt}  ,
\end{equation}
where $\mu_i$ are extra state variables whose evolution is governed by a linear time invariant system in Eqn. \ref{eqn:errorState}.
\begin{equation}
    \label{eqn:errorState}
    \scriptsize
    \frac{d\mathbf{\mu}}{dt} = \mathbf{\nu_A} X_p(t) + \mathbf{\nu_B} u(t), 
\end{equation}
where $\mathbf{\mu}$ is the vector containing $\mu_i$, $\mathbf{\nu_A}$ is an $n \times n$ matrix containing all error values $\nu_{i,k}$ associated with $A(t)$ matrix. Similarly $\mathbf{\nu_B}$ is an $n \times m$ matrix containing all error values  $\nu_{i,k}$ for the $B(t)$ matrix. 

Eqn. \ref{eqn:Prop} expresses PLIS solution as a congregation of: a) state space evolution assuming zero order hold of model coefficients, b) propagation of errors of the zero order hold assumption through the unperturbed system, and c) a new state space (Eqn. \ref{eqn:errorState}) defined by the errors in the model coefficients. Overall solution is a LTI system with an extended state space of $Z = X_P \bigcup \mathbf{\mu}$ given by -
\begin{equation}
\label{eqn:Hstate}
\scriptsize
    \frac{dH}{dt} = A_{ex}H + B_{ex} u, H = \begin{bmatrix}
    X_p\\
    \mathbf{\mu}
    \end{bmatrix}, A_{ex} = \begin{bmatrix}
        A(\tau_j)+\mathbf{\nu_A} & A(\tau_j)\\
        \mathbf{\nu} & \Theta_{nn}
    \end{bmatrix}, B_{ex} = \begin{bmatrix}
    B(\tau_j) + \mathbf{\nu_B}\\
    \Theta_{nm}
    \end{bmatrix}
\end{equation}

 where $\Theta_{nn}$ is an $n \times n$ matrix of zeros and $\Theta_{nm}$ is an $n \times m$ matrix of zeros. $H$ is a $2n \times 1$ vector, $A_{ex}$ is a $2n \times 2n$ matrix, $B_{ex}$ is a $2n \times m$ matrix. By solving this LTI system we can determine the error and use the maximum error to obtain the sub-interval $[\tau_j, \tau_{j+1}]$.  

\noindent{\bf Estimate of $\nu_{i,k}$:} We assume the most conservative error in the zero order hold assumption where $\nu_{i,k}(t) = $ $[max_{\forall t\in [\tau_j,\tau_{j+1}]}$ ${\frac{dA_{i,k}(t)}{dt}}]t$, i.e. we take the maximum slope of $A_{i,k}(t)$ in the interval $[\tau_j,\tau_{j+1}]$ and assume that $\nu_{i,k}$ is linear. To ensure that the true error in zero order hold never exceeds this limit, we have to consider time sub-intervals such that $A_{i,k}(t)$ are monotonously increasing or decreasing. This can be achieved by examining the differential of each time varying model coefficient and setting the $q_{inv}$ to the minimum time required for the slope of any $A_{i,k}$ to change resulting in the following constant for $\nu_{i,k}$ 
\begin{equation}
\label{eqn:nu}
\scriptsize
 \nu_{i,k} = [max_{\forall t\in [\tau_j,\tau_{j+1}]}{\frac{dA_{i,k}(t)}{dt}}]q_{inv}. 
\end{equation}
\noindent{\bf Error bound and invariant time step estimation:}
\label{sec:simError}
Replacing $\nu_{i,k}$ in Eqn. \ref{eqn:Hstate} and solving for $H$ using the traditional Euler method gives the temporal evolution of $H$ in terms of $q_{inv}$. The difference between the first $n$ elements of $H$ and the estimate of $X_P$ by the PLIS gives an upper bound of simulation error. This is an upper bound because when selecting the approximation of $A_{i,k}$, we assume the maximum slope with which $A_{i,k}(t)$ varies over time $t$. By changing the $q_{int}$ the error can be modulated such that the error in trajectory is within $\epsilon_p$ and the error in trace is less than $\psi_p$.

\section{Evaluation}
We discuss evaluation metrics, the performance
results and comparison between the PLIS and ORACLE.

\noindent{\bf Artificial Pancreas (AP) System:} A continuous glucose monitor (CGM) senses glucose, a controller computes insulin delivery, which is executed through an infusion pump. The glucose insulin dynamics is given by the Bergman Minimal Model (BMM):
\begin{equation}
\label{eqn:BMM}
\scriptsize
   \begin{bmatrix}
       i\\
       i_s\\
       G
   \end{bmatrix} = A \begin{bmatrix}
       i\\
       i_s\\
       G
   \end{bmatrix} + B \begin{bmatrix}
       i_b\\
       0\\
       u_2
   \end{bmatrix},   A = \begin{bmatrix}
        -n & 0 & 0\\
        p_2 & -p_1 & 0\\
        0 & -G_b & -p3
    \end{bmatrix}, B = \begin{bmatrix}
        p_4\\
        0\\
        1/VoI
    \end{bmatrix}
\end{equation}
The input vector $u(t)$ consists of basal insulin level $i_b$ and the glucose appearance rate $u_2$. The state vector $X(t)$ has the blood insulin level $i$, the interstitial insulin level $i_s$, and the blood glucose level $G$. $p_1$, $p_2$, $p_3$, $p_4$, $n$, and $1/V_o I$ are all patient specific coefficients.

\begin{figure}
\center
\includegraphics[trim=0 0 0 0,width=0.75\columnwidth]{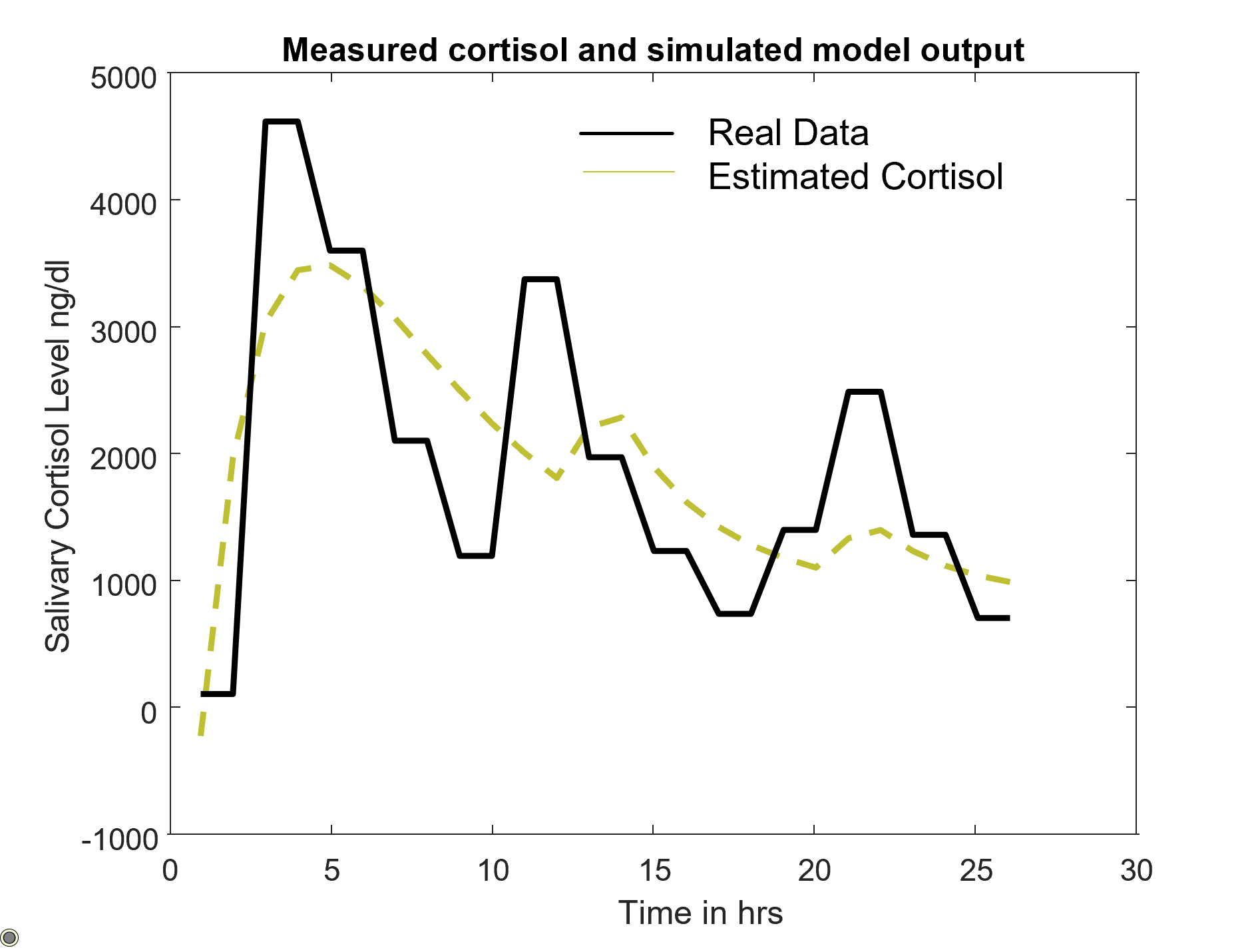} 
 \caption{Variation of cortisol level over time throughout the day of an individual with T1D under TSST.}
    \label{fig:Cort}
    \vspace{-0.1 in}
\end{figure}
\noindent{\bf Time variance:} The insulin sensitivity (SI) of a person varies with psychological stress throughout the day. SI is the parameter $p_3/p_4$. The psychological stress can be measured using the oral cortisol level of a person. SI has a negative linear correlation with cortisol~\cite{adam2010cortisol} which in turn varies over time (Fig. \ref{fig:Cort}). The time varying cortisol level $C(t)$ can be modeled using the following equation:
\begin{equation}
\label{eqn:Ct}
\scriptsize
    C(t) = \frac{K_p}{T_{p2}-T_{p1}}[e^{\frac{-t}{T_{p1}}} - e^{\frac{-t}{T_{p1}}}+ \frac{T_z}{T_d}[\frac{1}{T_{p1}}e^{\frac{-(t-T_d)}{T_{p1}}}-\frac{1}{T_{p2}}e^{\frac{-(t-T_d)}{T_{p2}}} ]]
\end{equation}
The SI is a linear function of the cortisol $C(t)$ and is given by a linear regression function 
\begin{equation}
\label{eqn:p3}
\scriptsize
    p_3(t) = p_4 \times (\eta C(t) + \beta),
\end{equation}
 where $\eta$ and $\beta$ are constants given in ~\cite{adam2010cortisol}. The time varying version of Eqn. \ref{eqn:BMM} is obtained by replacing $p_3$ by Eqn. \ref{eqn:p3}.

\noindent{\bf AP Dataset:} Through our collaboration with clinical partners at Mayo Clinic we have access to cortisol measurements (Fig. \ref{fig:Cort}) during a day of induced psychological stress through the Trier social stress test (TSST). The TSST was administered twice, once at 8 am and then again at 2:30 pm. Salivary cortisol level was measured at intervals of 30 mins. 12 subjects with Type 1 Diabetes were recruited. These subjects were using a Tandem Control IQ and MPC. Each subject was monitored for 3 weeks, with cortisol data collected only in one day. The parameters of Eqn \ref{eqn:Ct} were derived from the data with $T_d = 15 mins$. The computed poles and zeros are $T_{p1} = 2.5 hrs$, $T_{p2} = 5 hrs$, $T_z = 1.5 hrs$. The $K_p = 1.215 ng/dl $ value is obtained by fitting a linear regression model on the maximum value of cortisol in each cortisol wave in Fig. \ref{fig:Cort}.



\noindent{\bf AP control systems:} Three control strategies are used:

\noindent{\it a) Proportional integrative and derivative (PID) controller,} described in~\cite{messori2018individualized}. The PID gain values are obtained from the study in~\cite{messori2018individualized}.

\noindent{\it b) Model predictive control (MPC),} uses a model of the plant to predict future glucose values and then derives the appropriate insulin output to optimize an objective function. The model used in our MPC implementation is Eqn. \ref{eqn:BMM}. The prediction and control windows were set to 60 mins and 30 mins respectively as per implementation documentation in Control IQ AP system~\cite{brown2018first}.

\noindent{\it c) Optimal control with Bayesian meal prediction,} is a meal to meal controller that uses the linearized BMM and Linear Quadratic Gaussian (LQG) optimal control strategy to reach the set point before the next meal is taken. We modeled the meal patterns of each individual patient as large, medium and small meals. We utilized a Markov chain to represent the meal intake pattern of a patient and to predict the next meal size. The linearized patient model without bolus is then simulated to derive the largest possible CGM variation. This is then used to derive the set point for the current meal.

\noindent{\bf ORACLE simulator:} The oracle simulator uses the actual times of the real events in the dataset and simulates using an event-driven approach. In this approach, the exact timings of each event is known and the simulation is stopped. It is then reconfigured with the changes given by the HIL user, and the simulation is restarted. In between two events, the simulation uses the time varying version of Eqn. \ref{eqn:BMM}, and solves it using ODE45.

\noindent{\bf Koopman Simulator:} The Koopman simulator is very similar to the ORACLE simulator except that the plant physical dynamics are approximated with a higher order linear model. We used the CGM and insulin data from the AP dataset and executed the DMD reduced order linear system identification strategy to obtain a 13 order system. The Koopman operators for the 13 order system was obtained using the Koopman extracted code provided by the seminal publication~\cite{williams2015data} which connects DMD with Koopman theory. The DMD also gives the inverse measurement functions $M_f^{-1}$ that converts the 13 order system back to the 3 order system that directly correlates with the glucose dynamics shown in Eqn. \ref{eqn:BMM}.

\noindent{\bf PLIS Implementation:} The PLIS implementation uses $q_{inv}$ derived from Eqn. \ref{eqn:nu}, and the sim error bounding strategy described in Section \ref{sec:simError}. The only time varying parameter is the insulin sensitivity $p_3$, with the initial value obtained from~\cite{messori2018individualized}, which includes the values of all other parameters. The value of $q_{inv}$ is obtained for $\epsilon_p$ of [2\%, 5\% and 10\%] and the $\psi_p$ of [5\%, 10\%, and 15\%].

To determine the value of $q_{inv}$ we follow Algorithm \ref{alg:Algo}. For a given $\epsilon_p$ and $\psi_p$, we first choose a large $q_{inv}$. We then compute the maximum slope of $C(t)$ and compute the $A(\tau_j)$ and $B(\tau_j)$ matrices. We then simulate $H$ and $X_p$ following equations \ref{eqn:Hstate} and \ref{eqn:stateP}. The difference between the first $n$ elements of $H$ and $X_p$ is computed using the root mean square metric. If the maximum error is greater than $\epsilon_p$ then $q_{inv}$ is reduced by a value $d$. After looping through all time intervals in $[0, T]$, the overall error for the trace $T$ is computed and compared with $\psi_p$. Again if the error is greater than $\psi_p$, the $q_{inv}$ is reduced by $d$ and the simulation is rerun. The simulation is stopped when both the trajectory error and trace error is satisfied by $q_{inv}$.  

\begin{algorithm}
	\caption{$q_{inv}$ = Error\_Bounded\_Invariant\_Time($\epsilon_p$,$\psi_p$)}
	\scriptsize
\begin{algorithmic}[1]
		\State \textbf{input} $A$, $B$, and $C(t)$, WMN control algorithm,
		\State \textbf{output} Invariant Time Step $q_{inv}$
		\State Set an initial $q_{inv}$ and an initial value of $d$
            \While{$\exists \tau_j \forall j \in \{1 ldots T/q_{inv}\}$ }
            \State Compute the maximum of $\frac{dC(t)}{dt}$ from Equation \ref{eqn:Ct}
            \State Setup matrices $A(\tau_j)$ and $B(\tau_j)$
            \State Compute $H$ from Equation \ref{eqn:Hstate} and $X_p$ from Equation \ref{eqn:stateP}
            \State Compute root mean square error between the first $n$ elements of $H$ and $X_p$.
            \State Set $r^j_{max}$ = maximum rmse out of all $n$ rmses. 
            \If{$r_{max} > \epsilon_p$}
                \State $q_{inv} = q_{inv} - d$
            \Else
                \State remove $\tau_j$ from the set.
            \EndIf
            \EndWhile
            \State Total error $E = \sum_{\forall j}{r^j_max}$
            \If{$E > \psi_p$}
            \State $q_{inv} = q_{inv} - d$
            \State Rerun the while loop
            \Else
            \State Return $q_{inv}$ 
            \EndIf 
\end{algorithmic}
	\label{alg:Algo}
\end{algorithm}
\noindent{\bf Evaluation Metrics}

\noindent{\it a) Glycemic metrics:} We compare the three simulators in terms of Time in range (TIR) i.e. $70\leq CGM \leq 180$, Time above range (TAR) i.e. $CGM > 180$, and TIme below range (TBR) i.e. $CGM < 70$. 

\noindent{\it b) Optimality ratio:} $\rho = $ $\frac{(mean(G_P) + mean(I_P) + mean(({i_s}_P))}{(mean(G_o) + mean(I_o) + mean(({i_s}_o))}$, is defined as ratio of the mean values of the glucose, plasma insulin and interstitial insulin estimates for the PLIS or Koopman simulator to the mean values of the same metrics for ORACLE simulator. 

\noindent{\it c) Simulation speedup $S_p$:} is defined by the ratio of execution time of PLIS or Koopman simulator with respect to the ORACLE. 

\noindent{\bf Evaluation Results}

\begin{table}[t]
	\centering
	\caption{ Glycemic metrics.}
	\scriptsize
	\begin{tabular}{p{0.7 in}|p{0.7 in}|p{0.4 in}|p{0.4 in}|p{0.45 in}}
	 \toprule
		{Approach} & {Control Method } & {TIR (\%) } & { TAR (\%) } &{TBR (\%)}\\ \midrule
ORACLE & PID  & 69 [$\pm$ 17] & 29 [$\pm$ 18] & 2 [$\pm$ 1.3] \\
 & MPC & 76 [$\pm$ 14] & 20 [$\pm$ 15] & 4 [$\pm$ 3] \\
 & Bayesian & 83 [$\pm$ 21] & 11 [$\pm$ 8.2] & 6 [$\pm$ 4]\\\midrule
Koopman & PID  & 68 [$\pm$ 17] & 27 [$\pm$ 16] & 5 [$\pm$ 3] \\
 & MPC & 76 [$\pm$ 17] & 22 [$\pm$ 13] & 2 [$\pm$ 3] \\
 & Bayesian & 84 [$\pm$ 20] & 13 [$\pm$ 6.1] & 3 [$\pm$ 2]\\\midrule
 PLIS ($\epsilon_p =$ & PID  & 68 [$\pm$ 21] & 30 [$\pm$ 19] & 2 [$\pm$ 1.4] \\
 $3\%, \psi_p = 5\%$) & MPC & 76 [$\pm$ 17] & 21 [$\pm$ 11] & 3 [$\pm$ 2] \\
 & Bayesian & 82 [$\pm$ 24] & 13 [$\pm$ 5] & 5 [$\pm$ 2]\\\midrule
 PLIS ($\epsilon_p =$ & PID  & 71 [$\pm$ 27] & 36 [$\pm$ 17] & 3 [$\pm$ 1.7] \\
 $5\%, \psi_p = 10\%$) & MPC & 71 [$\pm$ 16] & 28 [$\pm$ 10] & 1 [$\pm$ 0.5] \\
 & Bayesian & 88 [$\pm$ 21] & 9 [$\pm$ 7] & 3 [$\pm$ 1]\\\midrule
 PLIS ($\epsilon_p =$ & PID  & 60 [$\pm$ 30] & 31 [$\pm$ 15] & 9 [$\pm$ 4.1] \\
 $10\%, \psi_p = 15\%$) & MPC & 82 [$\pm$ 19] & 16 [$\pm$ 6] & 2 [$\pm$ 0.7] \\
 & Bayesian & 74 [$\pm$ 14] & 20 [$\pm$ 14] & 6 [$\pm$ 5]\\
		\bottomrule
	\end{tabular}
	\label{tbl:Glycemic}
	\vspace{-0.15 in}
\end{table}

\noindent{\it Glycemic metrics under an WMN controller:}
Table \ref{tbl:Glycemic} shows that the ORACLE, Koopman and PLIS ($\epsilon_p = 3\%, \psi_p = 5\%$) all have very less difference in TIR, TAR, and TBR metrics. However, when the error margins of PLIS are relaxed then the WMN controller performance metrics deviate from the ORACLE.  We also see that both the Koopman and the PLIS also have similar relative glycemic metrics across the controllers. For Koopman and all the PLIS variations, the PID performed worse than the MPC, with the Bayesian approach showing the best TIR.

\begin{table}[t]
	\centering
	\caption{ Optimality and Speed up of simulation approaches.}
	\scriptsize
	\begin{tabular}{p{0.65 in}|p{0.65 in}|p{0.5 in}|p{0.55 in}}
	 \toprule
		{Approach} & {Control Method} & {Optimality $\rho$} & { Speedup $S_p$} \\\midrule
  Koopman & PID  & 1.1 & 1.2  \\
 & MPC & 1.15 & 1.1  \\
 & Bayesian & 1.2 & 1.05 \\\midrule
 PLIS ($\epsilon_p =$ & PID  & 1.12 & 2.1  \\
 $3\%, \psi_p = 5\%$) & MPC & 1.16  & 2,6  \\
 & Bayesian & 1.11 & 2.4 \\\midrule
 PLIS ($\epsilon_p =$ & PID  & 1.14 & 3.5  \\
 $5\%, \psi_p = 10\%$) & MPC & 1.21 & 4.1  \\
 & Bayesian & 1.17 & 3.7 \\\midrule
 PLIS ($\epsilon_p =$ & PID  & 1.23 & 7.4  \\
 $10\%, \psi_p = 15\%$) & MPC & 1.31 & 8.3  \\
 & Bayesian & 1.29 & 8.1 \\
		\bottomrule
	\end{tabular}
	\label{tbl:OPT}
	\vspace{-0.15 in}
\end{table}

\noindent{\it Optimality w.r.t ORACLE:}
The optimality metric (Table \ref{tbl:OPT}) shows that the Koopman simulator is closest to the ORACLE. The PLIS is also close to the ORACLE and the Koopman for the PID controller. However, for MPC the PLIS is further from the ORACLE. Moreover, as the error margin is relaxed, PLIS goes further than the ORACLE. 

\noindent{\it Speedup w.r.t ORACLE:} The PLIS has the better speedup with respect to ORACLE than Koopman. The PLIS has 1.7 to 8 times speedup with respect to ORACLE. As the error margin is relaxed, the speedup increases while optimality reduces. 

\section{Conclusions}
Time variance is seen in practical deployments of wireless mobile networks (WMN) especially when humans are involved in decision making (human in the loop) and also participate as a component of the plant (human in the plant). Time variance introduces non-linearities in the system dynamics. In this paper, we present a theoretical framework for evaluating the error bound for piecewise time invariant simulation. We provide an algorithm to bound simulation error within a bound while achieving speed up. We show its execution for a real world WMN simulation of the artificial pancreas mobile wireless control system. 
\section*{Acknowledgements}
We are thankful to Dr Yogish Kudva, and Dr. Ravinder jeet Kaur from Mayo Clinic Rochester for providing access to data. The work is partly funded by Helmsley Charitable Trust, DARPA AMP N6600120C4020 and Arizona New Economic Initiative PERFORM Science and Technology Center.
\bibliographystyle{ACM-Reference-Format}
\bibliography{abbrev}

\end{document}